\documentstyle[aps,prb,epsf]{revtex}

\begin{document}
\twocolumn[\hsize\textwidth\columnwidth\hsize\csname @twocolumnfalse\endcsname

\draft
\preprint{s/o/s, Feb. 1999}

\title
{\bf Theory of the Josephson effect in
superconductor /
one-dimensional electron gas / superconductor junction}
\author
{Yukio Tanaka}

\address{Department of Applied Physics, Nagoya University, 
Nagoya, 464-8603 Japan. }

\author
{Takashi Hirai, Koichi Kusakabe}

\address
{Graduate school of science and technology,
Niigata University,
Ikarashi, Niigata, 950-2181 Japan}

\author{Satoshi Kashiwaya}

\address{Electrotechnical Laboratory, Umezono, Tsukuba, Ibaraki,
305-8568, Japan.}

\date{\today}
\maketitle
\begin{abstract}
We present a theory for the Josephson effect in an
unconventional superconductor / one-dimensional electron gas /
unconventional superconductor ($s/o/s$) junction,
where the
Josephson current
is carried  by components
injected perpendicular to the interface.
When superconductors on both sides have triplet symmetries,
the Josephson current is enhanced
at low temperature due to the zero-energy states
formed near the interface.
Measuring Josephson current in this $s/o/s$ junction,
we can identify parity of the superconductor.
\end{abstract}

\vskip2pc]

Nowadays, novel interference effects of the
quasiparticle tunneling in unconventional superconductor
junctions, where pair potentials change sign on the Fermi surface,
have been paid much attention. 
\cite{Hu,Tanaka}
One of the remarkable features is the formation of
the zero-energy states (ZES)
localized near surfaces of 
unconventional superconductors \cite{Hu,Buchholtz,Hara}.
The ZES are detectable by tunneling spectroscopy
as conductance peaks.
Experimental observations of the ZES on surfaces of high-$T_{c}$
superconductors have been reported in several papers.
\cite{Geerk,Lesueur,Alff,Wei}
Motivated by these works,
general formulas for the Josephson current in
(even parity)
unconventional superconductors were presented
by taking account of the ZES.\cite{Tanaka2,Barash,Samanta,Riedel}
Calculated results show
several anomalous properties including the strong enhancement
of the Josephson current at low-temperature
under the influence of the ZES formation.
\par
Recently, Maeno, $et$ $al$.
discovered  superconductivity in Sr$_{2}$RuO$_{4}$,
where symmetry of the pair potential is believed to be
triplet.\cite{Maeno,Rice,Machida}.
In (odd parity) triplet superconductor
junctions, it is also expected
that the
Josephson current is enhanced by the
formation of the ZES
similarly to the even parity cases.\cite{Hara,Yamashiro}
Since the ZES formation is a universal phenomenon for any pair potential
with the sign change on the Fermi surface
irrespective of parity of the pair potential,
it is not so easy to determine
the parity of the unconventional superconductor
using usual Josephson junctions.
\par
In order to distinguish odd parity superconductors from even parity ones,
we propose a new method using
a superconductor / one dimensional electron gas
(1DEG) /superconductor
($s/o/s$) junction.
Anomalous behaviors in the Josephson effect are expected only
for odd parity  superconductor in this junction
configuration.
This is because direction of quasiparticle injection,
which is a decisive factor for the formation of the ZES,
is restricted to be normal to the interface.
In this configuration, the appearance of the
ZES is governed by the parity of the superconductor,
as precisely discussed below.
Thus $s/o/s$ junction provides a simple strategy
to determine the parity of the superconductor.
\par
Recent rapid progress in the technology of
superconductor / semiconductor hybrid structure makes it possible
to fabricate and to study $s/o/s$ junctions.
Hence the way is promising enough.
Several theories have already been presented about the effect of
interaction in 1DEG on the Josephson effect
using superconductor / Luttinger liquid (LL) /
superconductor ($s/LL/s$) junctions.\cite{Fazio,Maslov}
In these works, however,
the superconductor is assumed to be BCS-type $s$-wave and
cases for unconventional superconductors
are not clarified yet. 
\par
In this paper, a formula of the Josephson current is presented
for $s/o/s$ junctions assuming that the 1DEG is non-interacting.
The Josephson current is shown to be
sensitive to the parity of the superconductor.
We further study the effect of interaction for the 1DEG
using the Tomonaga-Luttinger (TL) model.
A Josephson-current formula for
general $s/o/s$ junctions with normal boundary reflections
is obtained by generalizing the method by Maslov {\it et al.}, 
which again shows sensitivity of the current
to the parity of the superconductor.
\par
Let us consider a semi-infinite superconductor
with a flat interface at $x=0$ as shown in Fig.1. 
The effective potentials for injected and reflected 
quasiparticles with spin index $\sigma$ are given by 
$\Delta_{L\sigma}(\theta)$ and $\Delta_{L\sigma}(\pi-\theta)$, 
respectively.
In usual Josephson junctions,
ZES at a surface are formed if a condition
$\Delta_{L\sigma}(\theta)\Delta_{L\sigma}(\pi-\theta)<0$ is satisfied 
\cite{Hu,Buchholtz,Hara}.
On the other hand, as we stated above,
the most remarkable difference in
$s/o/s$ junctions from usual Josephson junctions
is that only the components of the current which flow
perpendicular to the interface ($\theta$=0)
contribute to the Josephson current.
For singlet superconductors,
since $\Delta_{L\sigma}(0)=\Delta_{L\sigma}(\pi)$,
the  condition for the ZES
is never satisfied. 
On the other hand, for triplet superconductors,
since $\Delta_{L\sigma}(0)=-\Delta_{L\sigma}(\pi)$
is satisfied, ZES are always expected \cite{Buchholtz}. 
This is the reason why we propose a $s/o/s$ junction
to distinguish the parity of the superconductor.
\vspace*{2mm}

\begin{figure}
\epsfxsize=55mm 
\centerline{\epsfbox{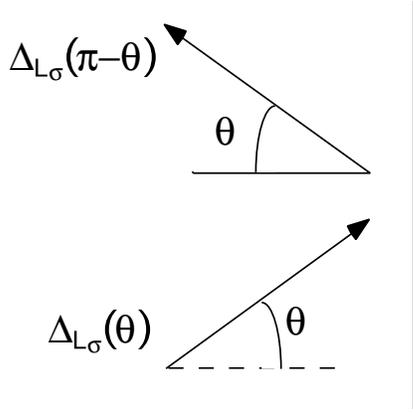}}
\vskip 4mm
\caption{
A schematic illustration for the formation of the
ZES at the surface of an unconventional superconductor.
}
\end{figure}
To perform the simplest model calculation, 
we consider $s/o/s$ junctions with perfectly flat interfaces
in the clean limit. In this model, the interface is perpendicular
to the $x$-axis and is located at $x=0$ and $x=d$ where
$d$ is the length of the 1DEG region.
In real junctions, insulator is inevitably
located between the superconductors and the 1DEG.
We model the insulator by a delta functions, namely
$H\delta (x)$ and $H\delta (x-d)$, where
$H$ denotes strength of the barrier.
We assume that the superconductors are two-dimensional.
The Fermi wave number $k_{F}$ and the effective mass $m$
are assumed to be equal in the left- and  right superconductors.
In the 1DEG, the magnitude of the Fermi wave number and
the effective mass are also chosen as $k_{F}$ and $m$, respectively.
In the following, we will calculate the
Josephson current in the $s/o/s$ junction shown in Fig.2.
For simplicity, the Cooper pair is assumed to be
formed by two electrons with antiparallel spins 
both for the singlet pairing and for the triplet pairing 
($S=1$, $S_{z}=0$). 
\vspace*{2mm}

\begin{figure}
\epsfxsize=70mm 
\centerline{\epsfbox{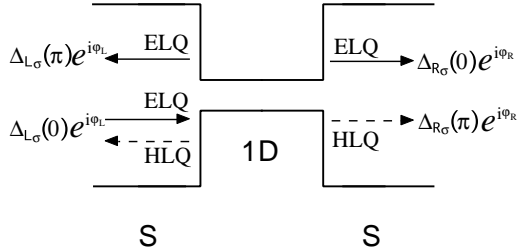}}
\vskip 4mm
\caption{
A schematic illustration of the
superconductor / 1DEG /  superconductor junction.
The effective pair potentials for an injected ELQ
(the reflected HLQ), the reflected ELQ, the transmitted ELQ
and the transmitted HLQ are
$\Delta_{L\sigma}(0)\exp(i\varphi_{L})$, 
$\Delta_{L\sigma}(\pi)$,
$\Delta_{R\sigma}(0)$, and $\Delta_{R\sigma}(\pi)$,
respectively.
}
\end{figure}

We first consider the case with non-interacting 1DEG.
In the framework of the quasi-classical approximation,
the effective pair potentials for the quasiparticles depend
on their directions of their motions.
We assume an electron-like quasiparticle  (ELQ) is  injected from the left.
The effective pair potentials for
the injected ELQ
[a reflected hole-like quasiparticle (HLQ)],
the reflected ELQ, the transmitted ELQ and the
transmitted HLQ are given by
$\Delta_{L\sigma}(0)\exp(i\varphi_{L})$,
$\Delta_{L\sigma}(\pi)\exp(i\varphi_{L})$,
$\Delta_{R\sigma}(0)\exp(i\varphi_{R})$,
and $\Delta_{R\sigma}(\pi)\exp(i\varphi_{R})$,
respectively (see Fig. 2).
The quantities $\varphi_{L}$ and $\varphi_{R}$ denote the
macroscopic phases, which are measured along the $x$-axis,
of the left and right superconductors, respectively.
The Josephson current through the junction is expressed
in terms of the coefficients of the Andreev reflection \cite{Andreev}
[$a_{\sigma}(\varphi)$] as
\begin{equation}
\label{1.0}
R_{N}I(\varphi)
=\frac{\pi k_{B}T}{ e \sigma_{T}}
\sum_{\omega_{n},\sigma}
\frac{\Delta_{L\sigma}(0)}{2\Omega_{n}}
[a_{\sigma}(\varphi) - a_{\sigma}(-\varphi)]
\end{equation}
with $\Omega_{n}
=\sqrt{\omega_{n}^{2} + \mid \Delta_{L\sigma}(0) \mid^{2}}$,
$\varphi=\varphi_{L}-\varphi_{R}$, and
$\omega_{n}=2\pi k_{B}T(n+ 1/2)$ with an integer, $n$
\cite{Furusaki}.
Conductance of the junction in the normal state
$\sigma_{T}$ is given by
\begin{eqnarray}
\label{1.1}
\sigma_{T}
&=&\frac{\sigma_{N}^{2}}
{ \{
1+(1-\sigma_{N})^{2}+
F(1-\sigma_{N})\}
} \\
F &=&
[2(2\sigma_{N}-1)\cos(2k_{F}d) +
4\sqrt{\sigma_{N}(1-\sigma_{N})} \sin(2k_{F}d)] \nonumber
\end{eqnarray}
with $\sigma_{N}=4/(4+Z^{2})$ and $Z=2mH/\hbar^{2}$.
Coefficients of the Andreev reflection are
obtained by solving the following equations,
\[
\Psi(x=0_{-})=\Psi(x=0_{+}), \ \
\Psi(x=d_{-})=\Psi(x=d_{+}),
\]
\[
\frac{d}{dx}\Psi(x) \mid_{x=0_{+}}
- \frac{d}{dx}\Psi(x) \mid_{x=0_{-}}
=\frac{2mH}{\hbar^{2}}\Psi(x) \mid_{x=0_{+}}
\]
\begin{equation}
\frac{d}{dx}\Psi(x) \mid_{x=d_{+}}
- \frac{d}{dx}\Psi(x) \mid_{x=d_{-}}
=\frac{2mH}{\hbar^{2}}\Psi(x) \mid_{x=d_{+}},
\end{equation}
where $\Psi(x)$ denotes the two component wave functions.
In the following, we will consider two cases;
(1)
singlet superconductor /  1DEG /
singlet superconductor ($ss/o/ss$) junction
[$\Delta_{L(R)\sigma}(0)
=\Delta_{L(R)\sigma}(\pi)=s \Delta_{0}$],
(2)
triplet superconductor / 1DEG /
triplet superconductor ($ts/o/ts$) junction
[$\Delta_{L(R)\sigma}(0)= \Delta_{0}$,
$\Delta_{L(R)\sigma}(\pi)=- \Delta_{0}$],
with $s=1$ ($s=-1$) for up (down) spin electron injection.
The Josephson current is expressed as  \par
\noindent
(1)$ss/o/ss$ junction case; \par
\begin{equation}
R_{N}I(\varphi)=
\frac{\pi k_{B}T}{e \sigma_{T}}
\sum_{\omega_{n}}
\frac{4\gamma \eta^{2} \sigma_{N}^{2} \sin \varphi}
{ \sigma_{N} \Lambda + (1-\sigma_{N})(1+\eta^{2})^{2}t }
\end{equation}
\noindent
(2)$ts/o/ts$ junction case; \par
\begin{equation}
R_{N}I(\varphi)=
\frac{\pi k_{B}T}{e\sigma_{T}}
\sum_{\omega_{n}}
\frac{4\gamma \eta^{2} \sigma_{N}^{2} \sin \varphi}
{ \sigma_{N}\Lambda + (1-\sigma_{N})(1-\eta^{2})^{2}t }
\end{equation}
\noindent
where
\[
\Lambda =
(1+\gamma^{2}\eta^{4}+2\gamma\eta^{2}\cos\varphi )
-(1-\sigma_{N})(\gamma^{2}+\eta^{4}+2\gamma\eta^{2}\cos\varphi),
\]
\[
\eta=\frac{\Delta_{0}}{\Omega_{n} + \omega_{n}},
\gamma=\exp[-2\mid \omega_{n} \mid d/\hbar v_{F}],
\]
\begin{equation}
t=1 + \gamma^{2} -\gamma(t_{s}\delta + \frac{1}{t_{s}\delta}), \
t_{s}=-\frac{2-iZ}{2+iZ}, \
\delta=\exp(2ik_{F}d).
\end{equation}
Temperature dependence of the maximum Josephson current $I_{C}(T)$
of $ss/o/ss$ and $ts/o/ts$ junctions is plotted
in Fig. 3. With increasing $Z$, magnitude of
$R_{N}I_{C}(T)$ for $ss/I/ss$ junction is reduced.
On the other hand, for $ts/o/ts$ junction,
it is enhanced oppositely with increasing $Z$.
The enhancement of the Josephson current for
larger $Z$ is due to the resonating current
through the ZES formed
near the interface. In  real junctions, an insulating barrier
inevitably exists near the interface.
Such a situation corresponds to the larger magnitude of $Z$
in our calculations.
The present result suggests that
we can distinguish the parity of the superconductor,
whether $I_{C}(T)$ shows an upturn curvature (triplet case)
or not (singlet case).
\vspace*{2mm}

\begin{figure}
\epsfxsize=60mm 
\centerline{\epsfbox{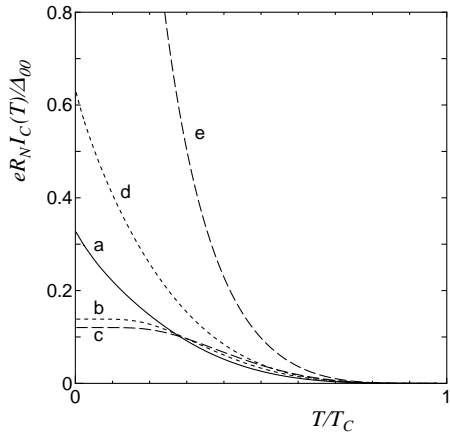}}
\vskip 4mm
\caption{
The maximum Josephson current $I_{C}(T)$
in $ss/o/ss$ junctions for (a) $Z=0$, (b) $Z=1$ and (c) $Z=5$
and that in $ts/o/ts$ junctions for
(d) $Z=1$ and (e) $Z=5$
with $dk_{F}=50$, $d/\xi=5$, and $\xi=\hbar v_{F}/\Delta_{0}$.
For $Z=0$, both junctions show the same magnitude of
$I_{C}(T)$. 
The $\Delta_{00}$ is the value of $\Delta_{0}$ at zero temperature, 
where the temperature dependence of $\Delta_{0}$ is assumed 
to obey the BCS relation. 
}
\end{figure}

Now, we consider an effect of the interaction in the 1DEG.
We derive the Josephson current formula for unconventional superconductors
with arbitrary barrier heights by taking account of both
the Andreev reflection and normal reflection at the interfaces.
The effect of interaction in 1DEG is introduced
following Maslov {\it et al.} using the TL-model.
\cite{Maslov,Haldane}
Basis for 1DEG is spanned by bound states formed in
the superconducting gap.
For simplicity, we consider here only the low temperature limit and
assume that relevant excitations determining the
Josephson current have energy, 
$\mid \varepsilon \mid \ll \Delta_0$.
Within these conditions,
a difference between $ss/LL/ss$ and $ts/LL/ts$ junctions
appears only in the following generalized boundary conditions
for the fermion field operators,
\begin{eqnarray}
\psi_{\pm,s} (x+2d) = \lambda \psi_{\pm,s} (x) \\
\psi_{+,s} (x) = s \psi^\dagger_{-,-s} (-x)
\end{eqnarray}
Here $\psi_{\pm,s}$ represents right-going (left-going) fermion
field with spin $s$. Extra phase factor 
$\lambda$, which is a function 
of $k_F$, $d$, $\sigma_N$, and $\varphi$,
coincide with the factor in Eqs. (16a), (16b) of Ref. [18],
when a $ss/LL/ss$ junction with only the Andreev refrection
at the boundary is considered.
Following the bosonization technique
for the open boundary conditions,
$\psi_{\pm ,s}$ can be represented by chiral
boson fields. \cite{Maslov,Fabrizio}
We see that only zero modes are affected by
the parity of the superconductor through the boundary
condition, (7), and
$\chi$ in the equation (28) of Ref. [18]
is replaced by a complicated function of $\varphi$ for general
situations considered here.
Explicit formulas for $\psi$ as well as $\lambda$ 
will be presented elsewhere.\cite{Hirai}
The current is obtained by
$I(\varphi )=-\frac{2ek_BT}{\hbar}
\frac{\partial}{\partial \varphi} \log Z(\varphi)$
where $Z(\varphi)$ is the partition function.
As Maslov {\it et al.} have claimed,
the Josephson current in the present limit is determined
by the zero mode (the topological excitations) and
non-zero modes do not contribute.\cite{Maslov}
Our general formula of the Josephson current 
for interacting 1DEG systems shows
essentially the same feature as non-interacting cases
in that $I(\varphi)$ is enhanced for
$ts/LL/ts$ compared with $ss/LL/ss$. \par
In this paper,
we propose a new method to identify the parity of a superconductor
using a $s/o/s$ junction.
We derive a formula for the Josephson current
assuming that the 1DEG is non-interacting.
Anomalous behavior in the Josephson effect is expected only
in triplet superconductor with odd parity.
This is because the direction of quasiparticle injection,
which is a decisive factor for the formation of ZES,
is selected to be normal to the interface \cite{Buchholtz}.
For the singlet superconductor with even parity,
the ZES never appear in the present geometry
as precisely discussed.
In the present calculation, the suppression of the
pair potential near the interface\cite{Rainer} is neglected.
Even if we take into account of this effect,
qualitative features in the upturn curvature due to the ZES 
at low temeratures will not  be changed, 
then the present results are still valid.\cite{Barash,Tanaka3} 
We have further studied the effect of interaction for the 1DEG 
using the TL model.
It is shown that the essential feature
is determined by the parity of the superconductor and
the influence of the interaction effect is not so important
within the TL model at the low temperature limit.
We will report detailed properties of general
$s/LL/s$ junctions in a forthcoming paper using a bosonization technique
with further
consideration of the inter-electron interaction \cite{Hirai}.
\par
\vspace{0.5cm}
This work has been supported by the Core Research for
Evolutional Science and Technology (CREST)
of the Japan Science and Technology Corporation
(JST) and a Grant-in-Aid for Scientific Research from the
Ministry of Education, Science, Sports and Culture.

\end{document}